\documentclass[twocolumn, preprintnumbers,amsmath,amssymb,balancelastpage]{revtex4}
\pdfoutput=1

\usepackage{float}
\usepackage{putexPRL}
\usepackage{graphicx}
\usepackage{dsfont}
\usepackage{amsmath}
\usepackage{array}
\usepackage{epstopdf}
\usepackage{enumerate}
\usepackage{youngtab}
\usepackage{tensor}
\usepackage{braket}
\usepackage[normalem]{ulem}
\usepackage{slashed}
\usepackage[aligntableaux=center]{ytableau}
\usepackage[utf8]{inputenc}
\usepackage[
      colorlinks=true,
      linkcolor=blue,
      urlcolor=blue,
      filecolor=black,
      citecolor=red,
      ]{hyperref}
\usepackage{braket}
\usepackage{mathtools}
\usepackage{rotating}
\usepackage{tabularx}
\newcolumntype{Y}{>{\centering\arraybackslash}X}
\newcolumntype{C}[1]{>{\centering\arraybackslash}p{#1}}
\usepackage{todonotes}
\usepackage{xcolor}
\definecolor{LightCyan}{rgb}{0.7,1,1}
\definecolor{Gray}{gray}{0.9}
\usepackage{xspace}

\newcommand {\be} {\begin {equation}}
\newcommand {\ee} {\end {equation}}

\newcommand {\bes} {\begin {equation*}}
\newcommand {\ees} {\end {equation*}}

\newcommand{\es}[2] {\begin{equation} \label{#1} \begin{split} #2 \end{split} \end{equation}}

\newcommand{\cO}{{\mathcal O}}

\newcommand{\beq}{\begin{equation}}
\newcommand{\eeq}{\end{equation}}

\def\ie{\begin{equation}\begin{aligned}}
\def\fe{\end{aligned}\end{equation}}

\def\<{\langle}
\def\>{\rangle}

\def\beg{\begin{equation}\begin{gathered}}
\def\eeg{\end{gathered}\end{equation}}

\def\bea{\begin{equation}\begin{aligned}}
\def\eea{\end{aligned}\end{equation}}

\begin{document}

\title{Bootstrapping Deconfined Quantum Tricriticality}

\author{Shai M.~Chester}         
\affiliation{
Jefferson Physical Laboratory, Harvard University, Cambridge, MA 02138, USA\\
Center of Mathematical Sciences and Applications, Harvard University, Cambridge, MA 02138, USA\\
Blackett Laboratory, Imperial College, Prince Consort Road, London, SW7 2AZ, U.K.
		}
\author{Ning Su}
\affiliation{Department of Physics, University of Pisa, I-56127 Pisa, Italy}

\begin{abstract}
The paradigmatic example of deconfined quantum criticality is the Neel-VBS phase transition. The continuum description of this transition is the $N=2$ case of the $CP^{N-1}$ model, which is a field theory of $N$ complex scalars in 3d coupled to an Abelian gauge field with $SU(N)\times U(1)$ global symmetry. Lattice studies and duality arguments suggest the global symmetry of the $CP^1$ model is enhanced to $SO(5)$. We perform a conformal bootstrap study of $SO(5)$ invariant fixed points with one relevant $SO(5)$ singlet operator, which would correspond to two relevant $SU(2)\times U(1)$ singlets, i.e. a tricritical point. We find that the bootstrap bounds are saturated by four different predictions from the large $N$ computation of monopole operator scaling dimensions, which were recently shown to be very accurate even for small $N$. This suggests that the Neel-VBS phase transition is described by this bootstrap bound, which predicts that the second relevant singlet has dimension $\approx 2.36$.
\end{abstract}

\maketitle
\nopagebreak

\section{Introduction}

Deconfined quantum critical points (DQCPs) are second order phase transitions between one phase with symmetry group $H$, and a second phase with group $H'$, where $H'$ is not a subgroup of $H$ \cite{doi:10.1126/science.1091806,2004PhRvB..70n4407S}. These phase transitions go beyond the standard Landau-Ginzburg transitions, such as the Wilson-Fisher fixed points, where $H$ would be a subgroup of $H'$. A striking feature of DQCPs is that they are described by continuum gauge theories in 2+1 dimensions whose fields are not associated with quasiparticles on either side of the transition, i.e. they are deconfined. Despite many years of work, however, the existence of the simplest DQCP remains controversial.

The paradigmatic example of a DQCP is the transition between the Neel and Valence Bond Solid (VBS) phases of quantum antiferromagnets on a 2d square lattice \cite{PhysRevLett.62.1694}, where the Neel phase breaks an $SU(2)$ symmetry, while the VBS phase breaks a different $U(1)$ symmetry. In the continuum limit, the Neel-VBS phase transition is described by the 3d $CP^1$ model \footnote{The equation of motion of $\lambda$ imposes the constraint $|\phi^i|^2=1$, which makes the theory equivalent to the non-linear sigma model with $CP^{1}$ target space \cite{coleman_1985}.} with Lagrangian
 \es{CPNActionI}{
\mathcal{L}=&\sum_{i=1}^2\left[|(\nabla_\mu-i A_\mu)\phi^i|^2+\lambda(|\phi^i|^2-1)\right] \,,
 }
 where $\phi_i$ are complex scalar fields, $A_\mu$ is an Abelian gauge field, and $\lambda$ is a real scalar. The $SU(2)$ symmetry rotates the $\phi_i$, while the $U(1)$ symmetry is generated by the current $\epsilon_{\mu\nu\rho}F^{\nu\rho}$, which is conserved due to the Bianchi identity \footnote{Including discrete symmetries, the full symmetry group is $SU(2)/\mathbb{Z}_2\times U(1)\times \mathbb{Z}_2^c$, where $\mathbb{Z}_2^c$ is charge conjugation. The discussion in this paper is not sensitive to these discrete symmetries, however, so we will not consider them in what follows.}. This theory is strongly coupled, so it is hard to determine if it actually flows to a conformal field theory (CFT), i.e. if it describes a second order phase transition. 
 
 \begin{table}
\begin{ruledtabular}
\begin{tabular}{c||l|l|l|l|l}
& $\Delta_v$ & $\Delta_t$ & $\Delta_{t_3}$ & $\Delta_{t_4}$& $\Delta_{s}$ \\ 
 \hline
 Bootstrap& $0.630^*$ & $1.519$ & $2.598$ & $3.884$ &$2.359$\\
 \hline
Large $N$&  $0.630$ & $1.497$ & $2.552$&$3.770$  &--\\
\hline
Lattice &  $0.630(3)$ & $1.5$ & --  &--&--\\
\hline
Fuzzy Sphere&  $0.584$ & $1.454$ & $2.565$&$3.885$  &2.845\\
\end{tabular}
\end{ruledtabular}
\caption{Comparison of scaling dimensions of the lowest dimension scalar operators in the singlet ($s$), vector ($v$), rank-2 ($t$), rank-3 ($t_3$), and rank-4 ($t_4$) of $SO(5)$, as determined from the bootstrap study here, the large $N$ expansion, the lattice study that claimed $SO(5)$ symmetry \cite{Nahum:2015vka}, and the recent weakly first order fuzzy sphere study for a certain value of their coupling \cite{Zhou:2023qfi}. The asterisk by $\Delta_v$ for bootstrap means we put it in to determine the others.
}
\label{results}
\end{table}
 
Instead of directly analyzing the continuum theory, lattice methods have been applied to various discrete quantum model that are believed to lie in the same universality class. These studies have produced a bewildering array of critical exponents \cite{2013PhRvL.111m7202B,2009PhRvB..80r0414L,PhysRevB.92.184413,2014PhRvB..89a4404S,2015PhRvB..91j4411P,PhysRevLett.98.227202,PhysRevLett.100.017203,PhysRevX.5.041048}, while others have claimed the transition is first order \cite{Song:2023wlg,2008arXiv0805.4334K,2008JSMTE..02..009J,PhysRevB.103.085104}. One notable lattice study suggested that the $SU(2)\times U(1)$ symmetry is enhanced to $SO(5)$ at the putative critical point \cite{Nahum:2015vka}, which was later attributed to possible quantum dualities \cite{Wang:2017txt}.

Most lattice studies so far have tried to find a fixed point by tuning one parameter, i.e. they assumed that there was one relevant operator uncharged under $SU(2)\times U(1)$ \footnote{More precisely, it was uncharged under whatever subgroup of these symmetries is preserved by the relevant lattice} \footnote{See however \cite{2020PhRvL.125y7204Z} for a recent lattice study that assumed tricriticality, but obtained very different scaling dimensions than us, and did not discuss $SO(5)$. See also \cite{chen2023phases,2019arXiv190410975L}, which suggested the possibility of a multicritical point.}. The $SO(5)$ symmetry enhancement would then imply there is no relevant $SO(5)$ singlet \footnote{As reviewed below, the $SU(2)\times U(1)$ singlet would become part of the rank-2 irrep of $SO(5)$.}, but this was also ruled out by the conformal bootstrap \cite{Li:2018lyb,Poland:2018epd} \footnote{A previous bootstrap study \cite{Nakayama:2016jhq} constrained the ability to find a critical point on lattices that preserve a $Z_r$ subgroup of $U(1)$, which is related to whether charge $q=r/2$ monopoles are relevant.}. This led some to propose the theory is described by a weakly first order phase transition, caused by the merger and annihilation \cite{PhysRevD.80.125005,Gorbenko:2018ncu} of the $CP^1$ model with a related tricritical model. This scenario received recent support by the study of a theory with microscopic $SO(5)$ symmetry using the fuzzy sphere method \cite{Zhou:2023qfi}.\footnote{A different fuzzy sphere calculation of this model suggested it was tricritical \cite{chen2023phases}, similar to the proposal in this work.}

%This assumption was ruled out by the conformal bootstrap study \cite{Nakayama:2016jhq}, which excluded the critical exponents predicted by \cite{2013PhRvL.111m7202B,2009PhRvB..80r0414L,PhysRevB.92.184413,2014PhRvB..89a4404S,2015PhRvB..91j4411P,PhysRevLett.98.227202,PhysRevLett.100.017203,PhysRevX.5.041048} under this assumption. Furthermore, the $SO(5)$ symmetry enhancement would imply there is no $SO(5)$ singlet \footnote{As reviewed below, the $SU(2)\times U(1)$ singlet would become part of the rank-2 irrep of $SO(5)$.}, and this was also ruled out by the bootstrap \cite{Li:2018lyb}. 

In this work, we will instead show evidence that the Neel-VBS phase transition is a tricritical fixed point, with two relevant singlets of $SU(2)\times U(1)$, or one relevant singlet of $SO(5)$. We use the fact that the scaling dimension $\Delta_q$ of local operators in the $CP^1$ model with charge $q\in\mathbb{Z}/2$ under the $U(1)$ symmetry, called monopole operators \cite{Murthy:1989ps}, can be computed to surprising accuracy using the large $N$ expansion in the related $CP^{N-1}$ model \cite{Metlitski:2008dw,Dyer:2015zha}. We review the evidence for this both from comparison to lattice studies for $q=1/2$ and various $N$ \cite{2013PhRvB..88v0408H,2012PhRvL.108m7201K}, as well as for $N=1$ and various $q$ \cite{Chester:2022wur} by comparing to the well-studied critical $O(2)$ model via particle-vortex duality \cite{Peskin:1977kp,PhysRevLett.47.1556}.

We then perform a conformal bootstrap study of $SO(5)$ invariant CFTs whose only relevant operators are the singlet $s$, rank-1 $v$, rank-2 $t$, and rank-3 $t_3$ scalars, as suggested by large $N$. The bootstrap rigorously bounds the space of allowed scaling dimensions of these operators, as well as of the irrelevant rank-4 scalar $t_4$. Physical theories often appear at the boundary of the allowed bootstrap region. In our case, by looking at the point on the boundary given by the large $N$ value of $\Delta_v$ and then maximizing $\Delta_t$, we can read off the values of all the other scaling dimensions. As shown in Table \ref{results}, our results for $\Delta_t$, $\Delta_{t_3}$ and $\Delta_{t_4}$ after imposing $\Delta_v$ all match the large $N$ estimates, and we also predict that $\Delta_s=2.36$. 

In Table \ref{results} we also compare our results to other studies that found $SO(5)$ symmetry: the original lattice study \cite{Nahum:2015vka}, and the recent fuzzy sphere paper \cite{Zhou:2023qfi} that starts from an $SO(5)$-invariant theory. While \cite{Nahum:2015vka} did not report a relevant $SO(5)$ singlet, since they assumed the $SU(2)\times U(1)$ theory was critical, the two scaling dimensions they did predict match ours \footnote{For $\Delta_v$, \cite{Nahum:2015vka} found very similar values for scaling dimensions of the Neel and VBS order parameters that combine to form $v$. We show the value of the Neel parameter in Table \ref{results}. }. Similarly, \cite{Zhou:2023qfi} argued that the theory is weakly first order, such that critical exponents depend  on the value of the coupling, but nonetheless we find that their results are similar to ours for a certain value of their coupling \footnote{In particular, for each system size they dial the coupling $V/U$ in their notation so that the stress tensor has dimension exactly three. We show the value given in their Table II which has scaling dimensions closest to our values.}, except that their singlet has slightly bigger dimension.

The rest of this paper is organized as follows.  In Section~\ref{sec:review}, we review properties of the $CP^{N-1}$ theory, including the large $N$ expansion of monopole operators, and the $SO(5)$ symmetry enhancement for $N=2$. In Section~\ref{sec:bootstrap}, we describe our bootstrap setup, how to numerically implement it using the Skydiving algorithm, and the resulting estimates for CFT data. We end with a discussion of our results in Section~\ref{sec:discussion}.

\section{The $CP^1$ model}\label{sec:review}

We now review the $CP^1$ model, first by generalizing to the $CP^{N-1}$ model at large $N$, and then by discussing the conjectured $SO(5)$ symmetry enhancement for $N=2$.

\subsection{The $CP^{N-1}$ model}
\label{sec:CPN}

We start with the Lagrangian of $N$ complex scalar fields $\phi_i$ coupled to an Abelian gauge field $A_\mu$ in 3d:
 \es{CPNAction}{
\hspace{-.1in}\mathcal{L}=&\sum_{i=1}^N\left[|(\nabla_\mu-i A_\mu)\phi^i|^2+m^2|\phi^i|^2\right]+u\Big[ \sum_{i=1}^N|\phi^i|^2\Big]^2 +\frac{F^2}{4e^2}  \,,
 }
 where $F_{\mu\nu}\equiv \partial_\mu A_\nu-\partial_\nu A_\mu$ is the field strength. At large $N$, we can tune $m^2=0$ to get a critical theory in the IR with $e,u\to\infty$, or tune both $m^2=u=0$ to get a tricritical theory in the IR with $e\to\infty$ \cite{PhysRevLett.32.292}. The critical theory is called the $CP^{N-1}$ model, since the $\phi^4$ interaction can be written in terms of a Hubbard-Stratonovich field $\lambda$ as in the $N=2$ case \eqref{CPNActionI}, in which case the theory is equivalent to a non-linear sigma model with $CP^{N-1}$ target space \cite{coleman_1985}. Both the critical and tricritical theories have an $SU(N)$ flavor symmetry that rotates the $\phi_i$, as well as a $U(1)$ topological symmetry whose current  $\epsilon_{\mu\nu\rho}F^{\nu\rho}$ is conserved due to the Bianchi identity \footnote{If we are careful about discrete groups, then the full global symmetry is $SU(N)/\mathbb{Z}_N\times U(1)\times \mathbb{Z}_2^c$, where $ \mathbb{Z}_2^c$ is charge conjugation.}. 
 
Local operators that transform under $SU(N)$ but not $U(1)$ can be constructed from the fields in the action $\phi_i$ and $A_\mu$, as well as $\lambda$ for the critical theory. The scaling dimension of these operators can be computed at large $N$ using standard Feynman diagrams \cite{PhysRevLett.32.292,Kaul_2008,Benvenuti:2018cwd,DIVECCHIA1981719,1996PhRvB..5411953I,Vasilev:1983uw}, but the large $N$ expansion is not very accurate for small $N$ \footnote{Some of these scaling dimensions have also been studied in the $d=4-\epsilon$ expansion, which is also not very accurate \cite{Moshe:2003xn,Folk1996OnTC,PhysRevB.100.134507}.}. 

\begin{table}
\begin{ruledtabular}
\begin{tabular}{c||l|l|l|l}
 $\Delta_{1/2}$ & $N=3$ & $N=4$ & $N=5$ & $N=6$ \\
 \hline 
Lattice&  $0.785$ & $0.865$ &$1.00(5)$ &$1.1(1)$ \\
 \hline
 Large $N$& $ 0.755$ & $ 0.880$ & $1.01$ &$1.13$\\
\end{tabular}
\end{ruledtabular}
\caption{Comparison of lowest charge monopole scaling dimension $\Delta_{1/2}$ between large $N$ and lattice studies for $N=3,4$ \cite{2013PhRvB..88v0408H} (JQ model) and $N=5,6$ \cite{2012PhRvL.108m7201K} ($J_1-J_2$ model).}
\label{Nbig}
\end{table}

\begin{table}
\begin{ruledtabular}
\begin{tabular}{c||l|l|l|l}
 $\Delta_q$& $q=1/2$ & $q=1$ & $q=3/2$ & $q=2$ \\
 \hline 
$O(2)$&  $.519130434$ & $1.23648971$ & $2.1086(3)$ & $3.11535(73)$ \\
 \hline
 Large $N$& $.50609$ & $1.1856$ & $2.0087$ & $2.9546$ \\
\end{tabular}
\end{ruledtabular}
\caption{Comparison of charge $q$ monopole scaling dimension $\Delta_{q}$ computed at large $N$ extrapolated to $N=1$, to values of the dual rank-$2q$ operators in the critical $O(2)$ model as computed from the conformal bootstrap in \cite{Chester:2019ifh}.
}
\label{kremp}
\end{table}

Local operators that transform under $U(1)$ with charge $q\in \mathbb{Z}/2$ are not built from fields in the action, but instead these monopole operators are defined as inserting magnetic flux $q=\frac{1}{4\pi}\int F$ \cite{Murthy:1989ps}. The lowest dimension monopoles are scalars and singlets under $SU(N)$ that we denote as $M_q$. Their scaling dimensions $\Delta_q$ are identified via the state-operator correspondence with the ground state energies in the Hilbert space on $S^2\times \mathbb{R}$ with $4\pi q$ magnetic flux through the $S^2$, which can be computed at large $N$ using a saddle point expansion \cite{Borokhov:2002cg,Metlitski:2008dw}. This calculation was carried out to subleading order in \cite{Dyer:2015zha}, and the results were found to be extremely accurate even at small $N$. For instance, for $q=1/2$ we compare the large $N$ estimates to lattice studies for $N=3,4,5,6$ \cite{2013PhRvB..88v0408H,2012PhRvL.108m7201K} in Table \ref{Nbig}. For $N=1$, the theory is dual to the $O(2)$ Wilson-Fisher fixed point \cite{Peskin:1977kp,PhysRevLett.47.1556}, so following \cite{Chester:2022wur} we can compare very precise bootstrap estimates of rank $2q$ operators in the critical $O(2)$ model to $M_q$ for low $q$ \footnote{Higher values of $q$ were also successfully matched in \cite{Chester:2022wur} by comparing to lattice data for the critical $O(2)$ model from \cite{Banerjee:2017fcx,Hasenbusch:2019jkj}. In \cite{Chester:2022wur}, the large $N,k$ expansion of $\Delta_q$ \cite{Chester:2017vdh,Chester:2021drl} was also shown to match the expected free theory dual \cite{Seiberg:2016gmd,Karch:2016sxi} when $N=k=1$, which is further evidence of the effectiveness of the the large $N$ expansion for monopoles.}, as we review in Table \ref{kremp}.

\subsection{$SO(5)$ symmetry enhancement}
\label{sec:SO5}

We now specialize to $N=2$ and discuss the conjectured enhancement of the $SU(2)\times U(1)$ global symmetry to $SO(5)$. In general, the rank-$2q$ symmetric traceless irrep of $SO(5)$ includes charge $\pm q,\pm(q-1),\dots$ irreps after decomposition to $SU(2)\times U(1)$. For instance, the vector ${\bf 5}$ of $SO(5)$ decomposes as
\es{vec}{
{\bf 5}\to {\bf 3}_0 \oplus {\bf1}_{\pm1/2}\,,
}
where ${\bf d}_q$ denotes the dimension $d$ (isospin $\frac{d-1}{2}$) irrep of $SU(2)$ with charge $q$ under $U(1)$, and ${\bf d}_{\pm q}$ means both ${\bf d}_q$ and ${\bf d}_{-q}$ appear. As discussed in \cite{Nahum:2015vka,Wang:2017txt}, we thus see that the VBS order parameter $M_{1/2}$ combines with the Neel order parameter $\phi^i\phi_j^\dagger$, which is the lowest dimension scalar operator in the adjoint of $SU(2)$, to form the lowest dimension vector operator $v$ of $SO(5)$. 

Similarly, the rank-2 ${\bf 14}$ of $SO(5)$ decomposes as 
\es{rank2}{
{\bf 14}\to {\bf 5}_0 \oplus {\bf 1}_0 \oplus {\bf3}_{\pm1/2} \oplus {\bf1}_{\pm1}\,.
}
Thus, the lowest dimension $SU(2)\times U(1)$ singlet scalar $\phi^i\phi^\dagger_i$ combines with $M_{1}$, the composite monopole operator $\phi^i\phi_j^\dagger M_{ 1/2}$, and a non-monopole operator with isospin 2 to form the lowest dimension rank-2 operator $t$ of $SO(5)$. This also implies that $\Delta_{\phi^i\phi^\dagger_i}=\Delta_1$, where $\Delta_1\approx1.497$ from large $N$. 

The rank-3 ${\bf 30}$ of $SO(5)$ decomposes as
\es{rank3}{
{\bf 30}\to {\bf 7}_0 \oplus {\bf 3}_0 \oplus {\bf5}_{\pm1/2} \oplus {\bf1}_{\pm1/2}  \oplus {\bf3}_{\pm1}  \oplus {\bf1}_{\pm3/2}\,,
}
which implies that $M_{3/2}$ joins with a $q=1/2$ scalar monopole as well as other operators to form the lowest dimension rank-3 operator $t_3$ of $SO(5)$. This $q=1/2$ scalar monopole cannot be the lowest dimension monopole $M_{1/2}$, because that was already used to form $v$, so it must be at least the second lowest dimension monopole $M'_{1/2}$ with $\Delta'_{1/2}=\Delta_{3/2}$. Since the large $N$ estimate gives $\Delta_{3/2}\approx 2.55$, this implies that the third lowest dimension $q=1/2$ scalar monopole that would be used to form the second lowest $SO(5)$ vector $v'$ according to \eqref{vec} must have an even bigger dimension, which strongly suggests that its irrelevant. An analogous argument suggests that the second lowest rank-2 $t_2'$ and rank-3 $t_3'$ must have dimensions bigger than $\Delta_2$ and $\Delta_{5/2}$ with large $N$ estimates $3.77$ and $5.12$ \cite{Dyer:2015zha}, respectively, which shows that $t$ and $t_3$ are the only relevant operators in their irreps. 

The rank-4 ${\bf 55}$ of $SO(5)$ decomposes as
\es{rank4}{
{\bf 55}\to& {\bf 9}_0 \oplus {\bf 5}_0\oplus {\bf 1}_0 \oplus {\bf7}_{\pm1/2} \oplus {\bf3}_{\pm1/2} \\
&\quad\, \oplus {\bf5}_{\pm1}  \oplus {\bf1}_{\pm1}\oplus {\bf3}_{\pm3/2}\oplus {\bf1}_{\pm2}\,,
}
which implies that $M_2$ joins with $M_1'$, a $SU(2)\times U(1)$ singlet, and other operators to form the lowest dimension rank-4 operator $t_4$ of $SO(5)$. Large $N$ gives $\Delta_2\approx3.77$, so this singlet is irrelevant, while a similar argument suggests that singlets that appear in the decomposition of higher rank $SO(5)$ operators are also irrelevant. If a second relevant $SU(2)\times U(1)$ singlet exists, then it must form the lowest dimension $SO(5)$ singlet scalar $s$. In terms of fields, this second lowest singlet would be some linear combination of $F^2$ and $\lambda^2$, which have dimension 4 at $N\to\infty$.

The mixed irrep ${\bf 35'}$ of $SO(5)$ decomposes as
\es{mixed}{
{\bf 35'}\to& {\bf 5}_0 \oplus {\bf 3}_0\oplus {\bf 1}_0 \oplus {\bf5}_{\pm1} \oplus {\bf5}_{\pm1/2}  \oplus {\bf3}_{\pm1/2}\,,
}
which includes an $SU(2)\times U(1)$ singlet. Since the lowest two such singlets were already used to form $t$ and $s$, this means the lowest dimension operator $\cO_{\bf 35'}$ must have at least $\Delta_{\cO_{\bf 35'}}\geq \Delta_s$, and probably much higher. The last irrep we consider is the ${\bf 35}$, which decomposes as 
\es{mixed2}{
{\bf 35}\to& {\bf 5}_0 \oplus {\bf 3}_0 \oplus {\bf 3}_0 \oplus {\bf5}_{\pm1/2} \oplus {\bf3}_{\pm1}  \oplus {\bf3}_{\pm1/2}\oplus {\bf1}_{\pm1/2}\,,
}
which includes a $q=1/2$ scalar monopole. Since the lowest two such operators were already used to form $v$ and $t_3$, this means the lowest dimension operator $\cO_{\bf 35}$ must have at least $\Delta_{\cO_{\bf 35}}\geq \Delta_{t_3}$, which suggests it is probably irrelevant.

\section{Numerical conformal bootstrap}\label{sec:bootstrap}

We will now describe our numerical bootstrap study of the $SO(5)$ invariant CFT.  

\subsection{Crossing equations}\label{cross}

Operators $\cO_{\bf r}(x)$ in the irrep ${\bf r}$ of $SO(5)$ can be written using fundamental indices $i=1,\dots,5$, and we will mostly be interested in rank-$q$ symmetric traceless tensors $\cO^{i_1\dots i_q}(x)$. Four-point functions of scalar operators $\varphi^{i_1\dots i_{q}}(x)$ can be expanded in the $s$-channel in terms of conformal blocks $g^{\Delta^-_{12},\Delta^-_{34}}_{\Delta,\ell}(u,v)$ \cite{Dolan:2003hv} as
\es{4point}{
\hspace{-.1in}&\left\langle  \varphi_{{\bf r}_1}(x_1)  \varphi_{\bf r_2}(x_2)   \varphi_{\bf r_3}(x_3)   \varphi_{\bf r_4}(x_4)  \right\rangle=  \frac{x^{\Delta^-_{12}}_{24} x^{\Delta^-_{34}}_{14}}{x^{\Delta^-_{12}}_{14} x_{13}^{\Delta^-_{34}} }      \\
\hspace{-.1in}&\times\frac{1} {x_{12}^{\Delta_{12}^+}x_{34}^{\Delta_{34}^+}} \sum_{\cO}\lambda^\cO_{\varphi_1\varphi_2}\lambda^\cO_{\varphi_3\varphi_4}T^{\bf r}_{\bf r_1\bf r_2\bf r_3\bf r_4}g^{\Delta^-_{12},\Delta^-_{34}}_{\Delta,\ell}(U,V),
}
 where $\Delta^\pm_{ij}\equiv\Delta_i\pm\Delta_j$, the conformal cross ratios $U,V$ are 
 \es{uv}{
   U \equiv \frac{{x_{12}^2x_{34}^2}}{{x_{13}^2x_{24}^2}},\qquad V \equiv \frac{{x_{14}^2x_{23}^2}}{{x_{13}^2x_{24}^2}}\,,
 }
and the operators $\cO$ that appear in both OPEs $\varphi_1\times\varphi_2$ and $\varphi_3\times\varphi_4$ have scaling dimension $\Delta$, spin $\ell$, and transform in an irrep $\bf r$ that appears in both the tensor products ${\bf r}_1\otimes{\bf r}_2$ and ${\bf r}_3\otimes{\bf r}_4$. If $\varphi_1=\varphi_2$ (or $\varphi_3=\varphi_4$), then Bose symmetry requires that $\cO$ have only even/odd $\ell$ for ${\bf r}$ in the symmetric/antisymmetric product of ${\bf r}_1\otimes{\bf r}_2$ (or ${\bf r}_3\otimes{\bf r}_4$). Equating this $s$-channel expansion with the $t$-channel expansion, given by swapping $\varphi_{{\bf r}_1}(x_1)$ and $\varphi_{{\bf r}_3}(x_3)$, gives the crossing equations
\es{crossing}{
0=& \sum_{\cO}\lambda^\cO_{\varphi_1\varphi_2}\lambda^\cO_{\varphi_3\varphi_4}T^{\bf r}_{\bf r_1\bf r_2\bf r_3\bf r_4}V^{\frac{\Delta_{23}^+}{2}}g^{\Delta^-_{12},\Delta^-_{34}}_{\Delta,\ell}(U,V)\\
&-\sum_{\cO}\lambda^\cO_{\varphi_3\varphi_2}\lambda^\cO_{\varphi_1\varphi_4}T^{\bf r}_{\bf r_3\bf r_2\bf r_1\bf r_4}U^{\frac{\Delta_{12}^+}{2}}g^{\Delta^-_{32},\Delta^-_{14}}_{\Delta,\ell}(V,U)\,,
}
which can be further decomposed into a finite set of equations as a function of $U,V$ using the explicit form of the tensor structure $T^{\bf r}_{\bf r_1\bf r_2\bf r_3\bf r_4}$. We consider correlators of the lowest dimension singlet $s$, vector $v$, and rank-2 $t$ scalar operators. In table \ref{configs} we list the 4-point functions of $s$, $v$, and $t$ that are allowed by $SO(5)$ symmetry and whose $s$ and $t$-channel configurations lead to independent crossing equations, along with the irreps and spins of the operators that appear in the OPE, and the number of crossing equations that they yield. These crossing equations were derived using the general $O(N)$ code \footnote{For these correlators, there is no difference between $SO(5)$ and $O(5)$.} from the project of \cite{He:2021sto}. We summarize our conventions for the tensor structures $T^{\bf r}_{\bf r_1\bf r_2\bf r_3\bf r_4}$ in Appendix \ref{sec:details}, while the explicit crossing vectors can be found in the attached \texttt{Mathematica} file. 

\begin{table}
\begin{ruledtabular}
\begin{tabular}{c|c|c|c}
 Correlator& $s$-channel & $t$-channel&Eqs\\
 \hline 
$\langle v v v v\rangle$&  ${\bf1_s}$, ${\bf10_a}$, ${\bf14_s}$&same&  3   \\
 \hline
 $\langle tttt\rangle$&  ${\bf1_s}$, ${\bf10_a}$, ${\bf14_s}$, ${\bf35'_s}$, ${\bf55_s}$, ${\bf81_a}$&same&  6   \\
 \hline
 $\langle t v t v\rangle$&  ${\bf5}$, ${\bf30}$, ${\bf35}$&same&  3   \\
 \hline
 $\langle tt v  v\rangle$&  ${\bf1_s}$, ${\bf10_a}$, ${\bf14_s}$ & ${\bf5}$, ${\bf30}$, ${\bf35}$ &  6   \\
 \hline
 $\langle ssss\rangle$&  ${\bf0_s}$& same &  1   \\
 \hline
 $\langle  v s v s\rangle$&  ${\bf5}$& same &  1   \\
 \hline
 $\langle tsts\rangle$&  ${\bf14}$ & same &  1   \\
 \hline
 $\langle ttss\rangle$&  ${\bf1_s}$ & ${\bf14}$ &  2   \\
 \hline
  $\langle v v ss\rangle$&  ${\bf1_s}$ & ${\bf5}$ &  2   \\
 \hline
   $\langle v s  v t\rangle$&  ${\bf5}$ & same &  1   \\
 \hline
 $\langle  v  v st\rangle$&  ${\bf14_s}$ &  ${\bf5}$ &  2   \\
 \hline
$\langle sttt\rangle$&  ${\bf14_s}$ &  same &  1   \\
\end{tabular}
\end{ruledtabular}
\caption{Four-point function configurations that give independent crossing equations under equating their $s$- and $t$-channels, where $\bf s/a$ denote that only even/odd spins appear.}
\label{configs}
\end{table}

\subsection{Numerical implementation}
\label{nav}

We bootstrap this system by truncating the 29 crossing equations listed in \ref{configs}, rephrasing them as a semidefinite program as in \cite{Poland:2011ey}, which crucially assumes unitarity, and then solving these constraints efficiently using \texttt{SDPB} \cite{Simmons-Duffin:2015qma}. We assume that $s$, $ v$, $t$, and $t_3$ are the only relevant scalar operators in any of the irreps that appear in Table \ref{configs}, which is supported by the large $N$ analysis of the previous section. For $\Delta_s$ and $\Delta_{\bf35'}$, we put put the slightly stronger gap 4 to improve numerical stability. We also impose that $s,v,t$ are unique \footnote{Since $t_3$ is not an external operator, we cannot impose uniqueness by scanning over its OPE coefficients.} by scanning over the four ratios of their OPE coefficients as in \cite{Chester:2019ifh}. The output of the bootstrap is an allowed region in the 8-dimensional space
$\big\{\Delta_ v\,, \Delta_s, \Delta_t \,, \Delta_{t_3}\,,
\frac{\lambda_{sss}}{\lambda_{ v v t}}\,, \frac{\lambda_{tts}}{\lambda_{ v v t}}\,, \frac{\lambda_{ v v s}}{\lambda_{ v v t}}\,, \frac{\lambda_{ttt}}{\lambda_{ v v t}}\big\}$.

Since this is a very large space, it would be computationally infeasible to map out the entire region, and we do not find any evidence that the allowed region is a small island. Instead, our hope is that the physical theory lies on the boundary of the allowed region, as was the case for bootstrap studies of many theories such as the critical $O(N)$ models \cite{ElShowk:2012ht,Kos:2013tga} and QED$_3$ with four fermions \cite{Chester:2016wrc}. Our strategy is to look at the point on the boundary of this 8-dimensional space given by imposing the value $\Delta_ v\approx .63$ that was predicted from the large $N$ analysis, maximizing $\Delta_t$, and then reading off the CFT data given by the approximate solution to the bootstrap equations at that boundary point \cite{ElShowk:2012ht}. We can do this using the recently developed Skydive method \cite{Liu:2023elz}, which is currently the most efficient way of finding the boundary of the allowed region. Even with this method, it takes over a week for the bootstrap to converge with bootstrap accuracy $\Lambda=19$ (See Appendix \ref{sec:details} for more details). The resulting scaling dimensions of all lowest dimension scalar operators up to rank 4, which is the highest rank we can access from our setup, are given in Table \ref{results}. Other CFT data, such as ratios of OPE coefficients, is summarized in Appendix \ref{sec:details}.

\section{Discussion}\label{sec:discussion}

In this work we showed that a point on the boundary of the allowed region of CFTs with $SO(5)$ symmetry corresponds to the large $N$ estimate of scaling dimensions of monopole operators in the $CP^1$ model. In particular, by inputting $\Delta_v$, we found that $\Delta_t$, $\Delta_{t_3}$, and $\Delta_{t_4}$ matched their large $N$ values. We also made a prediction for a relevant $SO(5)$ singlet $\Delta_s\approx2.36$, which implies that the $CP^1$ theory is a tricritical fixed point in terms of $SU(2)\times U(1)$ with relevant singlets: $\Delta_s$ and $\Delta_t\approx1.519$. It would be interesting to understand what happens to this putative SO(5) invariant CFT when it is perturbed by this scalar, and to determine the resulting phase diagram.\footnote{Some simulations have studied possible phase diagrams \cite{2019arXiv190410975L,chen2023phases}. However, it is still far from a conclusive answer.}

Curiously, aside from $\Delta_s$ our results are very similar to those of the recent fuzzy sphere model \cite{Zhou:2023qfi}, which claimed to observe a weakly first order transition. Recent lattice studies has also suggested the $CP^{N-1}$ model stops being critical below $N_c\approx7$ \cite{Song:2023wlg} or $4\leq N_c<10$ \cite{PhysRevB.103.085104}, perhaps because the critical and tricritical theories merge and go off into the complex plane. This is in some tension with the previous match between large $N$ and lattice studies for $N=3,4,5,6$  \cite{2013PhRvB..88v0408H,2012PhRvL.108m7201K}, as well as the fact that the $N=1$ theory is widely believed to be critical due to particle-vortex duality. A possible resolution is that these theories become tricritical below $N_c$. Indeed, a recent lattice study suggested that the $N=2$ theory is tricritical \cite{2020PhRvL.125y7204Z}, but reported some different critical exponents than us and did not discuss an enhanced $SO(5)$ \footnote{The authors have told us an upcoming work will discuss $SO(5)$ enhancement and tricriticality in more detail \cite{sandvik}.}. A possible resolution is that the critical and tricritical theories reemerge from the complex plane to become unitary CFTs below $N_c$. This might also explain why the large $N$ results for $\Delta_q$ in the $CP^{N-1}$ model, which is critical at large $N$, seem to match a tricritical theory at small $N$ as the analytic continuation in $N$ for $\Delta_q$ might switch between the critical and tricritical theories below $N_c$ \footnote{The question would remain of why the large $N$ analysis also matches the $N=1$ theory, which is critical. This could be due to the decoupling of the extra relevant singlet for some $1<N<2$. We thank Max Metlitski for discussion about this.}. It would be interesting if the bootstrap could be used to see the merger of the critical and tricritical theories as a function of real $N$, just as the merger and annihilation of the critical and tricritical 3-state Potts model was recently seen using the bootstrap as a function of dimension $2<d<3$ \cite{Chester:2022hzt}.

Looking ahead, we would like to improve our bootstrap study so that we can find a rigorous island around the large $N$ values. One way might be to bootstrap a system of correlators including the relevant rank-3 scalar $t_3$, as bootstrapping all relevant operators drastically improved bounds in other cases such as the critical $O(2)$ and $O(3)$ models \cite{Chester:2019ifh,PhysRevD.104.105013}. This would also give us access up to rank-6 operators, which could then be compared to the large $N$ predictions corresponding monopole operators. 

It would also be interesting to resolve a related tension between the lattice results for critical QED$_3$ with $N=2$ fermions \cite{Karthik:2016ppr,Karthik:2019mrr,Karthik:2015sgq}, whose $SU(2)\times U(1)$ symmetry was conjectured to enhance to $O(4)$ \cite{Hsin:2016blu,PhysRevB.92.220416,PhysRevLett.117.016802,Wang:2017txt}, and bootstrap bounds that ruled out these estimates \cite{Li:2021emd}. As in the $SO(5)$ case discussed here, the conflict with bootstrap can be avoided if we assume that $N=2$ QED$_3$ is tricritical. The large $N$ estimate for monopole operators has also been shown to be accurate at least for $N=4$, where there are independent boostrap results \cite{Chester:2016wrc,Albayrak:2021xtd}.

\section*{Acknowledgments}

We thank Cenke Xu, Max Metlitski, Ashwin Vishwanath, Subir Sachdev, Anders Sandvik, Ribhu Kaul, Silviu Pufu, Alessandro Vichi, Marten Reehorst, Zhijin Li, Yin-Chen He, and Slava Rychkov for useful conversations, Alessandro Vichi and Marten Reehorst for collaboration at an early stage of this project, and Max Metlitski for reviewing the manuscript. This project has received funding from the European Research Council (ERC) under the European Union’s Horizon 2020 research and innovation programme (grant agreement no. 758903), as well as from the Royal Society under the grant URF\textbackslash R1\textbackslash 221310. The authors would like to acknowledge the use of the Harvard cluster in carrying out this work. We thank Yin-chen He for support on computational resources. The computations in this paper were partially run on the Symmetry cluster of Perimeter institute. Research at Perimeter Institute is supported in part by the Government of Canada through the Department of Innovation, Science and Industry Canada and by the Province of Ontario through the Ministry of Colleges and Universities. The computations in this work were run on the Resnick High Performance Computing Center, a facility supported by Resnick Sustainability Institute at the California Institute of Technology. This work initiated at the GGI conference ``Boostrapping Nature'', and performed in part at Aspen Center for Physics, which is supported by National Science Foundation grant PHY-2210452.

\appendix

\section{Numerical bootstrap details}
\label{sec:details}

\begin{table}[t]
\begin{center}
\begin{tabular}{|l|l|}
\hline
%goal                                         & max $\Delta_t$   \\ \hline
precision                                    & 768                   \\ \hline
$\Lambda$                                    & 19                    \\ \hline
$\kappa$                                     & 14                    \\ \hline
spins                                        & $S_{19}$              \\ \hline
Initial Hessian                              & Identity              \\ \hline
$\texttt{dualityGapUpperLimit}$ for 1st SDP  & $10^{-6}$             \\ \hline
$\texttt{dualityGapUpperLimit}$ for the rest & None                  \\ \hline
$\texttt{centeringRThreshold}$               & $10^{-10}$            \\ \hline
%\hline
%total number of {\tt skydive}  calls: $d\xi, dp$               & 236                         \\ \hline
%total number of PDIP  iterations:   $d\xi$ & 580                            \\ \hline
%total number of Newton steps     & 816                       \\ \hline
\end{tabular}
\caption{\label{tab:param}Individual parameter setup for the computation to maximize $\Delta_t$.
}
\end{center}
\end{table}

The numerical calculations were performed using the \texttt{skydive} program \cite{Liu:2023elz} and \texttt{simpleboot} package \cite{simpleboot}. Some parameters are summarized in TABLE \ref{tab:param}, while for the remaining parameters we used the default values. The spin set is $S_{19}=\{0,..., 26\}\cup\{49,50\}$. A detailed description of the meaning of these parameters can be found in \cite{Liu:2023elz}. The most important parameters is $\Lambda$, as the bootstrap bounds monotonically get stronger as $\Lambda$ is increased.

%\section{Tensor structures}
%\label{tens}

The bootstrap equation for correlators involving $O(5)$ (which recall are identical to $SO(5)$ for these correlators) operators $v,s,t$ was derived during the project of \cite{He:2021sto} using the birdtracks method \cite{663b51c5-158f-325a-916b-03575665ab5f}. The code for producing the bootstrap equations, along with a brief explanation of the derivation can be found at \cite{PIcourseL2,NingL2}. Here we provide a brief summary of the convention for the OPE coefficients. The 4-pt tensor structure is given by
\begin{equation}
T^{\bf r}_{\bf r_1\bf r_2\bf r_3\bf r_4}= \frac{1}{\sqrt{\text{dim}(\bf r)}} \sum_{n} \left\{ \frac{n}{\bf r} \bigg| \frac{i, j}{\bf r_1, \bf r_2} \right\} \left\{ \frac{n}{\bf r} \bigg| \frac{k, l}{\bf r_3, \bf r_4} \right\}
\end{equation}
where $n,i,j,k,l$ are indices for $O(N)$ representation $\bf r,\bf r_1,\bf r_2,\bf r_3,\bf r_4$, and the symbol $\left\{ \frac{n}{\bf r} \bigg| \frac{k, l}{\bf r_3, \bf r_4} \right\}$ represents the Clebsch–Gordan (CG) coefficients, which satisfy the normalization $\sum_{a,b} \left\{ \frac{c}{\bf t} \bigg| \frac{a, b}{\bf r, \bf s} \right\} \left\{ \frac{c'}{\bf t} \bigg| \frac{a, b}{\bf r, \bf s} \right\}=\delta_{cc'}$. These conditions uniquely fix the convention of OPE coefficients.

We performed a run to maximize the $\Delta_t$. The final point we find is $\Delta_{s},\Delta_t,\Delta_{t_3}$ from Table \ref{results} and OPE coefficients
$\big\{\frac{\lambda_{sss}}{\lambda_{ v v t}}\,, \frac{\lambda_{tts}}{\lambda_{ v v t}}\,, \frac{\lambda_{ v v s}}{\lambda_{ v v t}}\,, \frac{\lambda_{ttt}}{\lambda_{ v v t}}\big\}=\{-0.28765, 1.28077, -0.70486, -0.31905\}$

As for performance detail of the computation, our computations were spread over multiple stages, where we explored various setups and gap assumptions. When we initiated the final computation that maximized the $\Delta_t$ at $\Lambda=19$, we already havd some knowledge of the expected value of the scaling dimensions and OPE coefficients. With our chosen initial point $\big\{\Delta_s,, \Delta_t \,, \Delta_{t_3}\,,
\frac{\lambda_{sss}}{\lambda_{ v v t}}\,, \frac{\lambda_{tts}}{\lambda_{ v v t}}\,, \frac{\lambda_{ v v s}}{\lambda_{ v v t}}\,, \frac{\lambda_{ttt}}{\lambda_{ v v t}}\big\}$=(2.1649, 1.4726, 2.4092, -0.31742, 1.2770, -0.71334, -0.44045), the run required 514 \texttt{skydive} calls. On a single node with 32 CPU cores, the run took approximately 7 days. For the computation of $\Delta_{t_4}$, we used the \texttt{spectrum.py} package.

\onecolumngrid
\vspace{1in}
\twocolumngrid

\bibliographystyle{ssg}
\bibliography{SO5_draft}

\end{document}